\documentstyle[fleqn,twoside,gc]{article}

\def\bce{\begin{center}}
\def\ece{\end{center}}
\def\beq{\begin{eqnarray}}
\def\eeq{\end{eqnarray}}
\def\ben{\begin{enumerate}}
\def\een{\end{enumerate}}

\def\brr{\begin{array}}
\def\err{\end{array}}

\heads{E. Elizalde}
      {Some uses of $\zeta-$regularization in quantum gravity and cosmology}

\begin{document}

\twocolumn[
\Arthead{6}{2000}{4 (24)}{1}{10}

\Title{SOME USES OF $\zeta-$REGULARIZATION IN  QUANTUM GRAVITY \\ 
 AND COSMOLOGY\foom 1}  

\Author{E. Elizalde\foom 2}{Instituto de Ciencias del Espacio (CSIC)\, 
\& IEEC, \, Edifici Nexus, Gran Capit\`{a} 2-4, 08034 Barcelona, Spain}

\Abstract{This is a short guide to some uses of the zeta-function
regularization procedure as a a basic mathematical tool for
quantum field theory in curved space-time (as is the case of
Nambu-Jona-Lasinio models), in quantum gravity models (in
different dimensions), and also in cosmology, where it appears
e.g. in the calculation of possible `contributions' to the
cosmological constant coming through manifestations of the vacuum
energy density. Part of this research was carried out in fruitful 
and enjoyable collaboration with people from Tomsk  State 
Pedagogical University.}


]  
\foox 1  {Paper for the special issue of {\sl Gravitation and Cosmology}
devoted to {\it Quantum Gravity, Unified Models and Strings}, to mark the 100th 
anniversary of Tomsk State Pedagogical University.}
\email 2 {elizalde@ieec.fcr.es, elizalde@math.mit.edu \\ \hspace*{4mm}
http://www.ieec.fcr.es/recerca/cme/eli.html}

\section{The method of zeta-function regularization}

Hawking  introduced this method \cite{hawk1} as a basic
 tool for the regularization of infinities in QFT
in a curved spacetime \cite{ramond1,bd1,bos1}. The idea is
the following \cite{hais1}. One could try to tame Quantum Gravity
 using the
canonical approach, by defining an arrow of time and
working on the space-like hypersurfaces perpendicular to it, with
equal time commutation relations. Reasons against
this: (i) there are many topologies of the
space-time manifold that are not a product {\bf R}$\times M_3$;
(ii) such non-product topologies are sometimes very interesting; 
(iii) what does it mean `equal time' in the
presence of Heisenberg's uncertainty principle?

One thus turns naturally towards the path-integral approach:
 \begin{eqnarray} <g_2,
\phi_2, S_2 | g_1, \phi_1, S_1> = \int {\cal D} [g, \phi] \
e^{iI[g, \phi]}, \end{eqnarray} where $g_j$ denotes the spacetime
metric, $\phi_j$ are matter fields,
 $S_j$ general spacetime surfaces ($S_j=M_j \cup \partial M_j$), $\cal D$ a measure over all
 possible `paths' leading from the $j=1$ to the $j=2$ values of
 the intervening magnitudes, and $I$ is the action:
\begin{eqnarray} I=\frac{1}{16\pi G} \int (R-2\Lambda) \sqrt{-g} \, d^4x +\int
L_m\sqrt{-g} \, d^4x, \end{eqnarray} $R$ being the curvature,
$\Lambda$ the cosmological constant, $g$ the determinant of the
metric, and $L_m$ the Lagrangian of the matter fields.
Stationarity of $I$ under the boundary conditions \begin{eqnarray}
\left. \delta g \right|_{\partial M} =0, \qquad \left.
\vec{n}\cdot \vec{\partial} \delta g \right|_{\partial M}
=0,\end{eqnarray} leads to Einstein's equations:
\begin{eqnarray} R_{ab} -\frac{1}{2} g_{ab} R+\Lambda g_{ab} =
8\pi G T_{ab}, \end{eqnarray} $T_{ab}$ being the energy-momentum
tensor of the matter fields, namely, \begin{eqnarray} T_{ab}=
\frac{1}{2\, \sqrt{-g}} \frac{\delta L_m}{\delta g^{ab}}.
\end{eqnarray} The path-integral
formalism  provides a way to deal
`perturbatively' with QFT in curved spacetime backgrounds \cite{bd1}. 
First, through a rotation in the complex plane one defines an Euclidean
action:
\begin{eqnarray} iI \longrightarrow -\hat{I}.\end{eqnarray} One can also
easily introduce the finite temperature formalism by the 
substitution $t_2-t_1 =i\beta$, which yields the partition
function
\begin{eqnarray} Z=\sum_n e^{-\beta E_n}. \end{eqnarray} If one
now adheres to the principle that the Feynman propagator is
obtained as the limit for $\beta \rightarrow \infty$ of the
thermal propagator, we have shown, some time ago \cite{ke1}, that
the usual principal-part prescription in the zeta-function
regularization method (to be described below) is actually {\it
not} needed any more, since it can in fact be replaced by this
more general principle.

Next comes the stationary phase approach (also called one-loop, or
WKB), for calculating the path integral, which consists in
expanding around a fixed background:
\begin{eqnarray} g=g_0 +\bar{g}, \qquad \phi =\phi_0 +\bar{\phi},
\end{eqnarray} what leads to the following expansion in the Euclidean
metric:
\begin{eqnarray} \hat{I}[g,\phi] =
\hat{I}[g_0,\phi_0]+I_2[\bar{g},\bar{\phi}] + \cdots
\end{eqnarray} This is most suitably expressed in terms of determinants (for bosonic, resp.
fermionic fields) of the kind (here $A,B$  are  the
relevant (pseudo-)differential operators in the corresponding
Lagrangian):
\begin{eqnarray} \Delta_\phi = \det \left( \frac{1}{2\pi \mu^2}
A\right)^{-1}, \quad \Delta_\psi = \det \left( \frac{1}{2 \mu^2}
B\right). \end{eqnarray}

\subsection{The zeta function of a $\Psi$DO and its associated
determinant}

\subsubsection{A pseudodifferential operator
($\Psi$DO)}

A {\it pseudodifferential operator} $A$ of order $m$ on a manifold
$M_n$ is defined through its symbol $a(x,\xi)$, which is a
function  belonging to the  space $S^m(\mbox{\bf R}^n\times
\mbox{\bf R}^n)$ of $\mbox{\bf C}^\infty$ functions such that for
any pair of multi-indexs $\alpha, \beta$ there exists a constant
$C_{\alpha,\beta}$ so that
\begin{equation}
\left| \partial^\alpha_\xi \partial^\beta_x a(x,\xi) \right| \leq
 C_{\alpha,\beta} (1+|\xi|)^{m-|\alpha|}.
\end{equation}
The definition of $A$ is given, in the distribution sense, by
\begin{equation}
Af(x) = (2\pi)^{-n} \int e^{i<x,\xi>} a(x,\xi) \hat{f}(\xi) \,
d\xi,
\end{equation}
where $f$ is a smooth function, $f \in
 {\cal S}$  [remember that ${\cal S} = \left\{ f \in  C^\infty
(\mbox{\bf R}^n);  \mbox{sup}_x |x^\beta \partial^\alpha f(x) | <
\infty, \right.$  $\left. \forall \alpha, \beta \in \mbox{\bf
R}^n\right\}$], $ {\cal S}'$ being the space of tempered
distributions and $\hat{f}$ the Fourier transform of $f$.
 When $a(x,\xi)$ is a polynomial in $\xi$
 one gets a differential operator.
In general,  the order $m$ can be complex. The {\it symbol} of a
$\Psi$DO  has the form
\begin{eqnarray}
a (x,\xi)& =&a _m(x,\xi) +a _{m-1}(x,\xi) + \cdots \nonumber \\
&&+a _{m-j}(x,\xi) + \cdots, \label{spsd}
\end{eqnarray}
being $a _k(x,\xi) = b_k(x) \, \xi^k$.

  Pseudodifferential
operators are useful tools \cite{eecmp1,eejcam1,eejpa01}, both in
mathematics and in physics.
 They were crucial for the proof of the
uniqueness of the Cauchy problem  and also for the proof of the
Atiyah-Singer index formula. In quantum field theory they appear
in any
 analytical continuation process (as
complex powers of differential operators, like the Laplacian). And
they constitute nowadays the basic starting point of any rigorous
formulation of quantum field theory through microlocalization, a
concept that is considered to be
 the most important step towards the understanding
of linear partial differential equations since the invention of
distributions.

\subsubsection{The zeta function} Let $A$ a positive-definite
elliptic $\Psi$DO of positive order $m \in \mbox{\bf R}$,
 acting on
the space of smooth sections of $E$, an $n$-dimensional vector
bundle over $M$, a  closed $n$-dimensional manifold. The {\it zeta
function} $\zeta_A$ is defined as \begin{eqnarray} \zeta_A (s) =
\mbox{tr}\ A^{-s} = \sum_j
 \lambda_j^{-s}, \qquad \mbox{Re}\ s>\frac{n}{m} \equiv s_0.
\end{eqnarray} where $s_0=$ dim$\,M/$ord$\,A$ is called the {\it abscissa of
convergence} of $\zeta_A(s)$ Under these conditions, it can be
proven that $\zeta_A(s)$ has a meromorphic continuation to  the
whole complex plane $\mbox{\bf C}$ (regular at $s=0$), provided
that the principal symbol of $A$ (that is $a_m(x,\xi)$) admits a
{\it spectral cut}: $ L_\theta = \left\{  \lambda \in \mbox{\bf
C};
 \mbox{Arg}\, \lambda =\theta,
\theta_1 < \theta < \theta_2\right\}$,   $\mbox{Spec}\, A \cap
L_\theta = \emptyset$ (Agmon-Nirnberg condition).
 The definition of $\zeta_A (s)$ depends on the
position of the cut $L_\theta$.
 The only possible singularities of $\zeta_A (s)$ are
{\it simple poles} at
$
s_k = (n-k)/m,  \  k=0,1,2,\ldots,n-1,n+1, \dots.
$
 M. Kontsevich and S. Vishik have managed to extend this
definition to the case when  $m \in \mbox{\bf C}$ (no spectral
 cut exists) \cite{kont95b}.

\subsubsection{The zeta determinant}

Let
 $A$ a $\Psi$DO  operator
with a  spectral decomposition: $ \{ \varphi_i, \lambda_i \}_{i\in
I}$, where $I$ is some set of indices. The definition of
determinant starts by trying to make sense of the product   $
\prod_{i\in I} \lambda_i$, which can be easily transformed into a
``sum'':
   $ \ln \prod_{i\in I} \lambda_i   = \
      \sum_{i\in I} \ln \lambda_i $. From  the
definition of the  zeta function of $A$:  $\zeta_A (s) =
    \sum_{i\in I} \lambda_i^{-s}$, by
 taking the derivative at $s=0$:   $\zeta_A ' (0) =  - \sum_{i\in I}
   \ln \lambda_i $, we arrive to the following definition of determinant
of $A$:
  $ \det_\zeta A = \exp \left[ -\zeta_A ' (0)
   \right] $.

A big amount of explicit formulas, useful for the calculation of
zeta functions and determinants of operators whose  spectrum is
known explicitly or implicitly (as roots of a spectral function
that incorporates the equation and the boundary conditions, think
e.g. of the roots of a Bessel function) are to be found, with all
sort of explanations, in our references
\cite{zb1,zb2,eecmp1,eejcam1,eejpa01,bek1}. Concerning the cases
of known spectrum, the most general situations we have been able
to consider can be summarized as follows. They are for zeta
functions of the form \cite{eejpa01}: \begin{eqnarray} \zeta_1 (s) & =&
\sum_{\vec{n} \in \mbox{\bf Z}^d} [Q(\vec{n}) +A(\vec{n})]^{-s},
\\  \zeta_2 (s) &=& \sum_{\vec{n} \in \mbox{\bf N}^d}
A(\vec{n})^{-s}, \end{eqnarray}  where $Q$ is a quadratic
non-negative form and $A$ a general affine form (they give rise to
 Epstein and  Barnes zeta functions,
respectively). I have also obtained explicit results (given in
terms of asymptotic series) for the much more involved cases where
the summation indices are `interchanged', namely: \begin{eqnarray}
\zeta_3 (s)& =& \sum_{\vec{n} \in \mbox{\bf N}^d} [Q(\vec{n})
+A(\vec{n})]^{-s}, \\ \zeta_4 (s)& =& \sum_{\vec{n} \in \mbox{\bf
Z}^d} A(\vec{n})^{-s}.\end{eqnarray}

\section{Four-fermion models in curved spacetimes}

Four-fermion models \cite{elo941,elos961} ---usually considered in
the $1/N$ expansion--- are interesting due to the fact that they
provide the opportunity to carry out an explicit, analytical study
of composite bound states and  dynamical chiral symmetry breaking.
At the same time, these theories ---and specially their
renormalizable 2d  and 3d  variants--- exhibit specific properties
which are similar to the basic behaviors of some realistic models
of particle physics. Moreover, this class of theories can be used
for the description of the standard model (SM) itself, or of some
particle physics phenomena in the SM.

Having in mind the applications of four-fermion models to the
early universe and, in particular, the chiral symmetry phase
transitions that take place under the action of the external
gravitational field, there has been some activity in the study of
four-fermion models in curved spacetime. The effective potential
of composite fermions in curved spacetime has been calculated in
different dimensions and dynamical chiral symmetry breaking,
fermionic mass generation and curvature-induced phase transitions
have been investigated in full detail. However, in most of these
cases only the linear curvature terms of the effective potential
had been taken into account. But it turns out in practice that it
is often necessary to consider precisely the strong curvature
effects to dynamical symmetry breaking. In fact we could see that
going beyond the linear-curvature approximation could lead to
qualitatively different results.

With the help of the zeta-function regularization techniques, we
investigated some 2d and 3d four-fermion models which were
renormalizable ---in the $1/N$ expan\-sion--- in a maximally
symmetric constant-curvature space (either of positive or of
negative curvature). The renormalized effective potential was
found for any value of the curvature and the possibility of
dynamical symmetry breaking in a curved spacetime was carefully
explored. Furthermore, the phase structure of the theory was also
described in detail.

 One should go to the original papers for details
\cite{elo941,elos961}. In particular, we calculated the effective
potential of composite fermions in the Gross-Neveu model, in the
spaces $S^2$ and $H^2$. The phase diagram in $S^2$ was constructed
and it was shown that for any value of the coupling constant there
exists a curvature above which chiral symmetry is restored. For
the case of $H^2$, we showed that chiral symmetry is always
broken. The asymptotic expansions of the effective potential were
given explicitly, both for small and for strong curvature. The
three-dimensional case was then studied. We considered two
different four-fermion models: one which exhibits a continuous
U(2) symmetry and another where we concentrated ourselves on two
discrete symmetries which happened never to be simultaneously
broken. We studied explicitly the dynamical P and Z$_2$ symmetry
breaking pattern in $H^3$ and $S^3$.

We started with the discussion of the Gross-Neveu model in the de
Sitter space. This model, although rather simple in its
conception, displays a quite rich structure, similar to that of
realistic four-dimensional theories
---as renormalizability, asymptotic freedom  and
dynamical chiral symmetry breaking. Some discussions of chiral
symmetry restoration in the Gross-Neveu model for different
external conditions (such as an electromagnetic field, non-zero
temperature or a change of the fermionic number density) had
appeared in the past (the influence of kink-antikink
configurations on the phase transitions was described too).

The study of the Gross-Neveu model in an external gravitational
field had been performed using the Schwinger method.
Unfortunately, the generalization of the Schwinger procedure to
curved spacetime is not free from ambiguities and this is why the
previous results  included some mistake, which we managed to
correct, by using a rigorous mathematical treatment of the
fermionic propagator in (constant curvature) spacetime. Our
starting point was the action
\begin{equation}
S= \int d^2x \, \sqrt{-g} \left[ \bar{\psi} i \gamma^\mu (x)
\nabla_\mu \psi + \frac{\lambda}{2N} (\bar{\psi}\psi)^2 \right],
\label{2.1}
\end{equation}
with $N$ the number of fermions, $\lambda$ the coupling constant,
$\gamma^\mu (x) = \gamma^a e^\mu_a (x)$, with $\gamma^a$ the
ordinary Dirac matrix in flat space, and $\nabla_\mu$  the
covariant derivative. By introducing the auxiliary field $\sigma$,
this could be rewritten  as
\begin{equation}
S= \int d^2x \, \sqrt{-g} \left[ \bar{\psi} i \gamma^\mu (x)
\nabla_\mu \psi - \frac{N}{2\lambda} \sigma^2 - \sigma
\bar{\psi}\psi\right],
\label{2.2}
\end{equation}
with $\sigma = - \frac{\lambda}{N} \bar{\psi}\psi$. Furthermore,
we went to Euclidean space.

There is no place for further details.  Using the powerful zeta-function
method, the renormalized effective potential was found for any
value of the curvature and its asymptotic expansion was  given
explicitly, both for small and for strong curvature. The influence
of gravity on the dynamical symmetry breaking pattern of some U(2)
flavor-like and discrete symmetries was described in detail. The
phase diagram in $S^2$ was constructed and it was shown that, for
any value of the coupling constant, a curvature existed above
which chiral symmetry was restored. For the case of $H^2$,
 it was always broken. In three dimensions,
in the case of positive curvature, $S^3$, it was seen that
curvature could induce a second-order phase transition.
 For $H^3$ the configuration
given by the auxiliary fields equated to zero was not a solution
of the gap equation. The physical relevance of the results was
discussed. In particular, we could see explicitly that the effect
of a negative curvature is similar to that of the presence of a
 magnetic field.

For the two-dimensional Gross-Neveu model on $S^2$ ---where the
chiral symmetry is a discrete one--- we showed the possibility of
chiral symmetry breaking and of fermion mass generation. Note that
the curvature of a two-dimensional de Sitter space acts here as
some external parameter (like temperature) which induces the
chiral symmetry phase transition. In this sense, and owing to the
fact that we treated curved spacetime {\it exactly}, the de Sitter
space could not be considered to be some fluctuation over flat
spacetime.

In the case of positive curvature, $S^3$, we  checked  that the
scenario given in the framework of the small curvature expansion
(e.g., fluctuations over flat space) changes dramatically when
gravity is treated exactly, what could certainly
 be useful in cosmological applications.

\section{Uses in dilaton gravity}

Two-dimensional dilaton gravity interacting with a four-fermion
model and scalars was investigated in a series of papers of our
collaboration (see, e.g., \cite{eno941,eo931,eonp931} to mention 
only a few). The one-loop covariant
effective action for 2D dilaton gravity with Majorana spinors
(including the four-fermion interaction) was obtained, and the
technical problems which appeared in any attempt at generalizing
such calculations to the case of the most general four-fermion
model described by Dirac fermions were properly discussed. A
solution to these problems was found, based on its reduction to
the Majorana spinor case. The general covariant effective action
for 2D dilaton gravity with the four-fermion model described by
Dirac spinors was given. The one-loop renormalization of dilaton
gravity with Majorana spinors was carried out and the specific
conditions for multiplicative renormalizability were found. A
comparison with the same theory but with a classical gravitational
field was also done.

Different approaches to the quantization of 2D dilaton gravities
(mainly, string inspired models) had been discussed already. We
obtained the covariant effective action corresponding to a very
general multiplicatively renormalizable
 model of 2D gravity with matter. Its action was of the following
form
\begin{eqnarray}
&& \hspace*{-8mm} S = - \int d^2x \sqrt{g} \left[ \frac{1}{2} 
Z( \Phi) g^{\mu \nu}
\partial_\mu \Phi \partial_\nu \Phi + C (\Phi ) R\right.\nonumber 
\\&& \hspace*{2mm} - \frac{i}{2}
q ( \Phi) \bar{\psi}_a \gamma^\lambda \partial_\lambda
\psi_a+ b(\Phi) \left(
\bar{\psi}_a  N_{ab} \psi_b \right)^2
 \nonumber \\ && \left.\hspace*{2mm}  -\frac{1}{2} f (\Phi)
g^{\mu\nu} \partial_\mu \chi_i
\partial_\nu
\chi_i + V(\Phi, \chi) \right]. \label{21}
\end{eqnarray}
It thus included a dilaton field, $\Phi$, $n$ Majorana fermions
$\psi_a$ interacting quartically via a symmetric constant matrix
$N_{ab}$, and $m$ real scalars $\chi_i$. We  also considered the
much more difficult case in which the action contained 2D Dirac
fermions (notice that we chose the matter to interact with the
dilaton via {\it arbitrary} functions).

This action described and generalized many well-known dilaton
models, e.g. the  bosonic string effective action
\begin{eqnarray}
&&\hspace*{-10mm}Z(\Phi )= 8 e^{-2\Phi}, \ C(\Phi) = 
 e^{-2\Phi}, \ V(\Phi)  =4
\lambda^2  e^{-2\Phi}, \nonumber \\ && \hspace*{6mm}
 q(\Phi) =  b(\Phi) = 0, \  f(\Phi) =1.
\end{eqnarray}
In the absence of matter our action for
\begin{equation}
Z=0, \ \ \ \ C(\Phi) =  \Phi, \ \ \ \ V(\Phi)  = \Lambda \Phi,
\end{equation}
coincided with the Jackiw-Teitelboim action. One could also add
gauge fields to the matter sector.

We  constructed the covariant effective action of the theory (\ref{21}),
studied its one-loop renormalization and discussed some thereby
connected issues. We described,  in full detail, the calculation of
the one-loop covariant effective action in 2D dilaton gravity with
Majorana spinors, what was the first example of such a kind of
calculation in two dimensions, as has been recognized latter by
many authors in their mentions to our paper. The inclusion of
scalars was also discussed there. We proceeded with the
computation of the covariant effective action of dilaton, scalars
and Majorana spinors for quantum systems in classical spacetime.
We discussed the technical problems which appeared in the derivation
of the covariant effective action in 2D dilaton gravity with Dirac
fermions: the solution of these problems was found, via reduction
of the system to the case of the theory of quantum dilaton gravity
with Majorana spinors. The one-loop renormalization of quantum
dilaton gravity with Majorana spinors was discussed too. The
conditions of multiplicative renormalizability were specified and
some examples of multiplicatively renormalizable dilaton
potentials were explictly obtained, as starting
point to discuss 2D quantum dilaton-fermion cosmology. There is no place here
to describe the usefulness of the zeta-function regularization procedure 
in all these developments and the reader is addressed to the original 
references, as quoted above (for a review of recent developments in 
this direction see \cite{no01}, and for  related result with possible 
cosmological application \cite{eno1}).

\section{Cosmological uses}

\subsection{The zero point energy}

If $H$ is now the Hamiltonian corresponding to a physical, quantum
system, the zero point energy is given by \begin{eqnarray} <0|
H|0>, \end{eqnarray} where $|0>$ is the vacuum state. In general,
after normal ordering we'll have: \begin{eqnarray} H= \left( n+
\frac{1}{2} \right) \, \lambda_n \, a_n\, a_n^\dagger,
\end{eqnarray} and this yields for the vacuum energy: \begin{eqnarray} <0|
H|0>= \frac{\hbar c}{2} \sum_n \lambda_n. \end{eqnarray} (I won't
normally keep track of the $\hbar$'s and $c$'s that will be set
equal to 1.) The physical meaning of this energy was the object of
a very long controversy, involving many first-rate physicists,
until the late Heindrik Casimir gave the explanation (over fifty
years ago) that is widely accepted nowadays, and that's the reason
why the zero-point energy is usually associated with his name.

The expression above acquires a very important meaning as soon as
one compares different settings, e.g., one where some sort of
boundary conditions are imposed to the vacuum (e.g., a pair of
parallel plates, infinitely conducting, in the vacuum
corresponding to the electromagnetic field) with another situation
where the boundary conditions (the plates) are absent (they have
been sent to infinity). The difference yields a physically
observable energy.

In general the sums appearing here are all divergent. They give
rise to the most primitive, but physically meaningful, examples of
zeta function regularization one can think of. In fact, according
to the definitions above:
\begin{eqnarray} <0| H|0> =\frac{1}{2} \zeta_H(-1).
\end{eqnarray}

It is important to notice that the zero-point energy is something
one always has to keep in mind when considering any sort of
quantum effect. Its contribution can be in some cases negligible,
even by several orders of magnitude (as seems to be the case with
sonoluminiscence effects), but it can be of a few percent (as in
some laser cavity effects), or even of some $10-30 \%$ as in the
case of several wetting phenomena of alcali surfaces by Helium.
Not to speak of the specifically devised experiments, where it may
account for the full result.

In the case of the calculation of the value of the cosmological
constant, it is immediate to see from the
expressions considered before that: \begin{eqnarray} <0|
T_{\mu\nu}|0> =\frac{\Lambda}{8\pi G} +\frac{1}{2 \, V} \sum_n
\lambda_n,
\end{eqnarray} where $V$ is the volume of the space manifold and
the second term as a whole is the vacuum energy density
corresponding to the quantum field (or fields) we are considering.
Unless the first term (the cosmological constant), the vacuum
energy density is not a constant (it goes as $a^{-4}$, $a$ being a
typical cosmological length). However, this does not prevent the
mixing of the two contributions when one considers, e.g., `the
presently observed value of the cosmological constant'. What
we have calculated is the second contribution for a scalar
field of very low mass.

\subsection{A simple model}

 Consider the space-time to be of one of the following
types: {\bf R}$\times${\bf T}$^p\times${\bf T}$^q$, {\bf
R}$\times${\bf T}$^p\times${\bf S}$^q$, $\ldots$, which are
actually plausible models for the space-time topology. A
(nowadays) free scalar field pervading the universe will satisfy
\begin{eqnarray} (-\Box +\xi R +M^2) \phi =0, \end{eqnarray} restricted
by the appropriate boundary conditions (e.g., periodic, in the
first case considered). We shall call $\rho_\phi$ the contribution
to $\rho_V$ from this field \begin{eqnarray}
 \rho_\phi =\frac{1}{2V} \sum_i \frac{\lambda_i}{\mu} =\frac{1}{2V}
\sum_{\mbox{\bf k}} \frac{1}{\mu} \left(k^2 +M^2\right)^{1/2},
\label{c2} \end{eqnarray} 
where the sums $\sum_i$ and
$\sum_{\mbox{\bf k}}$ are generalized ones (most common case: a
multidimensional series together with a multidimensional integral)
and $\mu$ is the usual mass-dimensional parameter to render the
eigenvalues adimensional (we take $\hbar =c =1$ and shall insert
the dimensionfull units only at the end of the calculation). The
mass $M$ of the field will be here considered to be arbitrarily
small and will be kept different from zero. This is nice, both
for computational reasons as well as for physical ones, since a
very tiny mass for the field can never be excluded.

 After going through some lengthy calculations that use the power of
zeta-regularization, we reach the conclusion that coincidence
with the observational value for the cosmological constant is
obtained for the contribution of a massless scalar field,
$\rho_\phi$, for $p$ large compactified dimensions and $q=p+1$
small compactified dimensions, $p=0,\ldots,3$, and this for values
of the small compactification length, $b$, of the order of 100 to
1000 times the Planck length $l_P$ (what is actually a very
reasonable conclusion, according also to other approaches). To be
noticed is the fact that full agreement is obtained only for cases
where there is exactly one small compactified dimension in excess
of the number of large compactified dimensions.
 $p$ and $q$ refer to the compactified dimensions
only, but there may be other, non-compactifed dimensions (exactly
$3-p$ in the case of the `large' ones). 
In particular, the cases of pure spherical
compactification and of mixed toroidal (for small magnitudes) and
spherical (for big ones) compactification can be treated in this
way and yield results in the same order of magnitude range. Both
these cases correspond to (observational) isotropic spatial
geometries. Also to be remarked again is the non-triviality of
these calculations, when carried out exactly, 
 what is apparent from the use of the generalized
Chowla-Selberg formula. Simple power counting is absolutely unable
to provide the correct order of magnitude of the results.

 The most precise fits
with the observational value of the cosmological constant are
obtained for $b$ between $b=100 \, l_P$ and $b=1000 \, l_P$, with
(1,2) and (2,3) compactified (large,small) dimensions, respectively. There
is in fact no tuning of a `free parameter' here.
 All them
correspond to a marginally closed universe, in full agreement too
with other completely independent analysis of the observational
data \cite{car,perl,ries,ries2}.

\Acknow{Thanks are given to my collaborators Andrei Bytsenko, 
S. Naftulin and
  Sergei Odintsov  for stimulating discussions
and a long and fruitful collaboration on these subjects. This
investigations have been supported by DGI/SGPI (Spain), project
BFM2000-0810, and by CIRIT (Generalitat de Catalunya), contract
1999SGR-00257.}

\end{document}